# Ultrafast melting and recovery of collective order in the excitonic insulator $Ta_2NiSe_5$

Hope M. Bretscher[1,†], Paolo Andrich[1,†,*], Prachi Telang[2], Anupam Singh[2], Luminita Harnagea[2], A. K. Sood[3], and Akshay Rao[1,*]

[1]*Cavendish Laboratory, University of Cambridge, Cambridge CB3 0HE, United Kingdom*

[2]*Department of Physics, Indian Institute of Science Education and Research, Pune, Maharashtra 411008, India*

[3]*Department of Physics, Indian Institute of Science, Bangalore, Karnataka 560012, India*

[†] These authors contributed equally

[*] Correspondence should be addressed to P. A. (pa343@cam.ac.uk) or A. R. (ar525@cam.ac.uk)

**Abstract:**

**The layered chalcogenide $Ta_2NiSe_5$ has been proposed to host an excitonic condensate in its ground state, a phase that could offer a unique platform to study and manipulate many-body states at room temperature. However, identifying the dominant microscopic contribution to the observed spontaneous symmetry breaking remains challenging, perpetuating the debate over the ground state properties. Here, using broadband ultrafast spectroscopy we investigate the out-of-equilibrium dynamics of $Ta_2NiSe_5$ and demonstrate that the transient reflectivity in the near-infrared range is connected to the system's low-energy physics. We track the status of the ordered phase using this optical signature, establishing that high-fluence photoexcitations can suppress this order. From the sub-50 fs quenching timescale and the behaviour of the photoinduced coherent phonon modes, we conclude that electronic correlations provide a decisive contribution to the excitonic order formation. Our results pave the way towards the ultrafast control of an exciton condensate at room temperature.**

## Introduction

Ultrafast optical spectroscopy has emerged in recent years as a crucial technique for the study and control of correlated electron phases in solid state materials[1–14] due to its few-femtosecond time resolution, ability to investigate phenomena on a rather wide energy scale, and superior signal-to-noise ratio[15]. This technique allows one to separate structural, spin, and charge contributions to the emergence of these systems' exotic behaviour by studying the timescales required to relax from an out-of-equilibrium state. This information is critical for engineering and manipulating these quantum phases[16]. For example, in the study of the properties of superconductors (SCs), ultrafast spectroscopy has been essential to uncover the microscopic origin of Cooper pair condensation[14].

The proposed excitonic insulating (EI) phase of matter shares numerous similarities in its theoretical treatment with the superconducting phase[17,18]. Below a critical temperature ($T_c$), an excitonic insulator undergoes a spontaneous symmetry breaking resulting in the formation of a macroscopic population of condensed electron-hole pairs in its ground state. Concurrently, a bandgap opens at the Fermi level, which mirrors the order parameter of the collective phase[19]. Interestingly, as a result of the strong Coulombic nature of the fermion-pair binding, this phase can appear at or above room temperature, providing a unique platform for the investigation of many-body effects. Additionally, EIs could manifest superfluidity and share features with other strongly correlated materials. One example is vanadium dioxide, where an insulating state emerges below a certain temperature as a result, at least partially, of electronic correlations. The insulator-to-metal photoinduced transition observed in this material has been proposed as a platform to obtain ultrafast Bragg switches[2].

Here, we investigate the properties of $Ta_2NiSe_5$, a layered van der Waals compound believed to be an EI[17,18] with $T_c$~328 K and an excitonic bandgap of ~150 meV at cryogenic temperatures. The stabilisation of the bound electron-hole pairs in this material has been attributed to its quasi-1D nature (see inset of Fig. 1b), with alternating chains of tantalum and nickel atoms in each of the stacked atomic planes[20]. A number of studies that support the hypothesis of an EI phase in $Ta_2NiSe_5$ have emerged in the past few years[21–26], and very recently the first detection of collective excitations in this system was reported[27] and explained theoretically[28,29]. However, the existence of a simultaneous structural phase transition at $T_c$[20] still calls into question the effective contribution of electronic correlations to the low-temperature behaviour of this material, and further efforts are required to directly study the properties of the EI order parameter and to develop a complete microscopic description of the interactions that govern this phase.

In this work, through a combination of temperature dependent, fluence dependent, and multi-pulse ultrafast broadband spectroscopy, we address these open questions. First, we demonstrate that the optical response of $Ta_2NiSe_5$ in the near infrared (NIR) range maps that of an order parameter. This result does not in itself provide information on the nature of the ordered phase, but rather means that we can use this response to track the out-of-equilibrium evolution of the system. In particular, we establish that a strong photoexcitation can completely suppress the ordered phase at room temperature. This result is confirmed through the study of the phononic degrees of freedom, which also manifest the response expected at high temperature. Using a two-pump technique, we can show that the evolution of the phonon response occurs on the same femtosecond timescale of the ultrafast electronic response. Notably, the timescale of the order melting and its dependence on the excitation fluence allows us to conclude that electronic correlations play a key role in driving the formation of the EI phase. Finally, above $T_c$, we observe a transient optical signal typical of systems with an energy gap at the Fermi level. We ascribe this behaviour to the existence of an insulating state of preformed excitons, as recently theoretically proposed[30].

## Results and discussion

We begin by characterizing the transient optical response of $Ta_2NiSe_5$ using pump-probe spectroscopy at room temperature. Fig. 1a shows the transient reflectivity ($\Delta R/R$) signal of this material obtained using 12 fs, 1.8 eV, low fluence ($\Phi_{pump}$ = 50 µJ/cm$^2$) pump pulses and probed using broadband (550 – 1250 nm) pulses. We observe an incoherent component to the signal, which quickly grows after time zero and then decays exponentially, and that is associated to the excitation of quasi-particles across the bandgap. The oscillations are instead associated to coherent phonon modes, which are excited through resonant impulsively stimulated Raman scattering or displacive excitation processes[31]. Contrary to previous reports[23], the $\Delta R/R$ signal does not depend on the pump polarisation, but it significantly changes with the orientation of the probe polarisation (see Supplementary Information, Note 1). We determine that the response to a polarisation oriented parallel to the atomic chains is more directly related to status of the condensate (as suggested in previous equilibrium measurements[32,33]) and limit our measurements to this probe configuration. This behaviour, together with the wavelength dependence described below, is in close analogy with that reported for superconducting systems[34–36], and can be ascribed to the difference in the initial and final states for the probe-induced transitions as a function of polarisation and wavelength. A detailed knowledge of the interband selection rules, which is beyond the scope of this work, is required to further investigate this aspect.

To identify the signatures of the proposed phase transition, we investigate the temperature dependence of the material response. In Fig. 1b, we present the spectrum of $\Delta R/R$ averaged between 1 and 2 ps for a series of temperatures and $\Phi_{pump}$ = 50 µJ/cm$^2$. While the signal shows a negligible change in the visible range (particularly at the shortest wavelengths), a more significant effect appears in the NIR, where the $\Delta R/R$ even flips sign as the temperature is increased above $T_c$. This behaviour becomes even clearer when we plot $\Delta R/R$ from this wavelength region as a function of time (Fig. 1c). Here we see that the overall signal dynamics profoundly change as a function of temperature. At room temperature, the signal magnitude increases over ~300 fs and subsequently decays over a time scale of a few hundred femtoseconds, while the oscillations continue over a much longer timescale. As we approach $T_c$ the signal amplitude rapidly decreases until it is completely replaced by a component of opposite sign that does not significantly decay on the timescale of our measurements. In Fig. 1d, we plot the signal amplitude extracted from a fit of the data (see Supplementary Information, Note 2) as a function of temperature. Interestingly, the amplitude follows remarkably well the behaviour of an order parameter with a zero temperature gap of 150 meV[37] (dashed line), changes sign close to $T_c$, and remains almost constant as the temperature is further increased. This behaviour is different from that previously observed at shorter probe wavelengths[38] (see also Supplementary Information, Note 3). The exact temperature at which the signal crosses zero depends weakly on the probe wavelength and it is sensitive to its polarisation, which again reflects the details of the transition matrix elements involved. However, the overall behaviour is clear at all wavelengths in the NIR range (see Supplementary Information, Note 3). This result exposes a connection between the low-energy physics of the excitonic condensate and the high energy electronic transitions probed by NIR light, in analogy with observations reported in high-temperature superconducting systems and other strongly correlated electron materials[36,39–41]. This interpretation is supported by previously reported results on the equilibrium response as a function of temperature[42]. Indeed, we can recognise a feature in the equilibrium optical conductivity at ~1 eV that maps the behaviour of the low-energy gap. Therefore, we interpret the drastic change in $\Delta R/R$ as the result of a shift in the spectral weight of the NIR electronic transition into the energy range of the excitonic bandgap when we increase the temperature above $T_c$ (see Supplementary Information, Note 4 for a more detailed explanation).

Focusing now on the NIR wavelength range, we investigate the effect of changing $\Phi_{pump}$ on the system's response at room temperature (Fig. 2a). In the low fluence regime, as expected, a stronger excitation results in a larger (negative) amplitude of $\Delta R/R$. However, as $\Phi_{pump}$ is further increased, the signal is progressively suppressed, until it becomes positive at early times, and then decays until reaching the level of the low fluence signal in ~1.5 ps. A closer inspection of the signal rise (Fig. 2b) reveals that the $\Delta R/R$ negative transient progressively becomes shorter as $\Phi_{pump}$ is increased, and it is absent at the highest fluences.

The behaviour of the coherent phonon oscillations can be studied performing a Fourier Transform (FT) of the oscillatory component of the signal. At room temperature the FT shows the well-known phonon resonances at 1, 2, 2.9, and 3.6 THz (Fig. 2c). As previously reported[23,43], the 2 THz oscillation becomes strongly suppressed as we approach $T_c$ and is then almost completely quenched up to ~400 K[44]. In Fig. 2d, we present instead the FT as a function of $\Phi_{pump}$. We see that while the 1, 2.9, and 3.6 THz phonon modes have a rather straightforward monotonic behaviour, the 2 THz mode is strongly modified at high $\Phi_{pump}$, where its resonance is shifted towards lower frequencies and is significantly broadened. Interestingly, recent works have suggested that the 2 THz phonon is the most strongly coupled to the electronic bands closest to the EI gap, and therefore could be the most influenced as the system is perturbed[45].

From these results we develop a microscopic description of the material response. At temperatures below $T_c$, the pump pulse excites highly energetic QPs across the single particle excitation gap. At low values of $\Phi_{pump}$, the QP population promptly starts to thermalise through electron-electron scattering events. As a part of this process, further QP excitations can be created through a cascade mechanism in which the QPs transfer some of their kinetic energy to carriers in the valence band. The rise of the $\Delta R/R$ amplitude reflects this carrier multiplication process. Within a few hundred femtoseconds the QPs relax to the edge of the single particle excitation gap through electron-phonon scattering processes. Subsequently, the QPs recombine across the single particle excitation bandgap and reform bound electron-hole pairs by emitting gap-energy bosons (amplitude or Higgs mode of the excitonic condensate[46]). This relaxation process occurs in ~2 ps, after which only the oscillations associated to the coherent phonons remain visible. This behaviour is different from that observed in SCs, where at temperatures below $T_c$ a bottleneck mechanism involving the re-breaking of Cooper pairs by the gap-energy bosons prolongs the QPs population lifetime beyond the intrinsic recombination rate[47,48]. We believe that this difference is the result of a crucial distinction between SCs and EIs in the low-energy range of the excitation spectrum. While in SCs the phase or Nambu-Goldstone mode of the condensate is pushed at energies above the Higgs mode due to the Anderson-Higgs mechanism[49], this is not the case in $Ta_2NiSe_5$, where we recently detected the phase mode signature at energies below the excitonic bandgap[27]. The presence of this low energy mode provides a channel for the rapid relaxation of the amplitude mode into pairs of phase modes, in analogy with the relaxation of optical phonons into pairs of acoustic ones[50]. These low-energy modes can quickly propagate away from the photoexcited region in addition to progressively relaxing into even lower energy excitations. This mechanism can prevent the bottleneck effect seen in SCs and allows for a fast relaxation of the QPs.

This result is in sharp contrast with the behaviour observed for temperatures above $T_c$, where the signal relaxation is drastically slowed down. The absence of a significant decay (see Supplementary Information, Note 2 for additional information) suggests that the system transitions to a gapped phase at high temperatures, with an energy gap larger than the available bosonic excitations in the system. In this context, the QPs quickly relax to the bottom of the respective energy bands but their lifetime is then limited by the intrinsic electron-hole recombination time across the bandgap, which can be of the order of nanoseconds in insulating systems[51]. In agreement with recent theoretical[30,52] and experimental[21] work, we propose that this insulating state is characterised by the existence of preformed excitons but no long-range order. While the debate over the nature of the high temperature phase of $Ta_2NiSe_5$ is still ongoing[24,53–56], this result clearly points at the existence of a gapped phase.

Considering now the data in Fig. 2, we can conclude that in the presence of a strong photoexcitation, the system undergoes a transient collapse of the EI phase associated to a sign inversion of $\Delta R/R$ at short pump-probe delays. As expected, the threshold fluence required for this collapse becomes smaller as the temperature is increased towards $T_c$, reflecting the presence of a weaker order (see Supplementary Information, Note 6). We attribute the duration of the negative transient seen in Fig. 2b to the time required for the complete melting of the excitonic order. This allows us to conclude that this process becomes progressively faster as we increase $\Phi_{pump}$ until the full quench of the excitonic order happens faster than our time resolution (see Supplementary Information, Note 5 for more details). The femtosecond timescale of this process excludes the possibility of a thermally driven transition, and instead we can associate this fluence dependent, ultrashort rise time to the existence of key electronic contribution to the ordered phase[57,58]. In addition, the very rapid relaxation of the $\Delta R/R$ signal towards negative values, which occurs on the timescale of the quasi-particles relaxation processes, also hints at the electronic nature of the driving force that restores the excitonic condensate. We note that, at long delay times, the signal for high $\Phi_{pump}$ acquires a small positive baseline, which we associate to bolometric heating as concluded in previous works[38].

The previous conclusions are well supported by our study of the phonon modes. In particular, the shift in frequency and increased linewidth of the 2 THz mode can be associated to an ultrafast transition to the high temperature lattice response that the system would present in the absence of stable electron-hole pairs[44]. Alternatively, this could be the result of the suppression of long range phase coherence upon photoexcitation[10,59–61]. This result further justifies associating the sign reversal of $\Delta R/R$ to a melting of the ordered phase.

The pump-probe measurements are key in uncovering the existence of a photoinduced phase transition, as evidenced both in the incoherent and coherent response of the system, and in studying the timescale of the order suppression. To gain more information on how the out-of-equilibrium state evolves over time, we can instead utilise a three-pulse scheme, equivalent to a pump-probe measurement on the out-of-equilibrium state of the system[10]. In particular (see Fig.3a), we use a first laser pulse (Pump1, P1) of varying fluence ($\Phi_{P1}$) to prepare the system into a perturbed state. After a delay $\Delta t_{P1-P2}$, a second weak pulse (Pump2, P2, 50 µJ/cm$^2$) further excites the material before a final pulse (Pr) is used to probe the state of the system after $\Delta t_{P2-Pr}$ (again we focus on the response of the system in the NIR). By subtracting the signal obtained in the presence of P2 and that collected with only the P1 excitation (we identify this signal as $\Delta(\Delta R/R)$), we can isolate the effect of the weak excitation on the perturbed state. By changing $\Delta t_{P1-P2}$ we can therefore study how the excitations induced by the second pump evolve in time (for more details on the implementation of this technique see [62]).

To further clarify this technique, in Fig. 3b and 3c we show $\Delta(\Delta R/R)$ as a function of $\Delta t_{P1-P2}$ for low (26 µJ/cm$^2$) and high (529 µJ/cm$^2$) values of $\Phi_{P1}$ respectively. In the presence of low P1 fluence, the $\Delta t_{P2-Pr}$ traces are identical for all $\Delta t_{P1-P2}$ delays, (and are similar to those seen at low fluence in Fig. 2a). However, when $\Phi_{P1}$ is increased, $\Delta(\Delta R/R)$ becomes positive for short $\Delta t_{P1-P2}$ delays. Only after ~1 ps of $\Delta t_{P1-P2}$ delay do the kinetics recover the low-fluence lineshape. To better illustrate this, in Fig. 3d, we plot the amplitude of the $\Delta(\Delta R/R)$ for a series of $\Phi_{P1}$ (we extract the amplitude using a fit process analogous to that discussed in Supplementary Information, Note 2). We emphasize again that this signal shows us the effect of P2 on the out-of-equilibrium response induced solely by the P1 pulse. Considering the data in Fig. 2a and 2b, this effect can be easily anticipated for low and high values of $\Phi_{P1}$. In these conditions, we know that an additional excitation results in a more negative and more positive transient reflectivity response respectively. More complex is the interpretation of the data for medium values of $\Phi_{P1}$. Here, we expect the $\Delta(\Delta R/R)$ signal to become positive as soon as the additional photoexcitation induced by P2 reduces the negative transient reflectivity value. Indeed, in Fig. 3d we see that the amplitude of $\Delta(\Delta R/R)$ becomes positive for values of $\Phi_{P1}$ much lower than the fluences needed to reverse the sign of $\Delta R/R$ (see Fig. 2a). As a consequence of this interpretation, we note that we do not expect to observe a rise time for a positive $\Delta(\Delta R/R)$ signal in the same way that we do in Fig. 2b. For instance, for all the four highest fluences used there, we expect the $\Delta(\Delta R/R)$ signal to be positive even at the shortest $\Delta t_{P1-P2}$ values. In principle, there could be very specific values of $\Phi_{P1}$ for which the $\Delta(\Delta R/R)$ signal would be positive for short values of $\Delta t_{P1-P2}$ and $\Delta t_{P2-Pr}$, but this would not have the same meaning as a change of sign in the transient reflectivity.

We can now interpret the behaviour of the $\Delta(\Delta R/R)$ amplitude for $\Phi_{P1}$ = 106, 176, and 529 µJ/cm$^2$ as that associated to a progressively stronger suppression of the excitonic order (all these curves decay with a similar recovery time $\tau_{cond}$, see Supplementary Information, Note 7). For $\Phi_{P1}$ = 882 µJ/cm$^2$ instead the amplitude decay is clearly slowed down, which could signify the complete suppression of the condensate and the onset of bottleneck processes that delay the electron-hole recombination rate and drives the system to the gapped phase. The value that the signal amplitude reaches at long delays depends on the number of long-lived QPs that can absorb the incoming P2 pulse, and is additionally influenced by the bolometric contribution.

So far, we have focused on the incoherent component of the $\Delta(\Delta R/R)$ signal, which confirms the conclusions reached from the data in Fig. 2 and provides additional insights on the recovery processes of the perturbed excitonic condensate. The three pulse technique can give us further understanding of the behaviour of the perturbed system when we look at the coherent component of the signal. At each $\Delta t_{P1-P2}$ delay, we perform a FT of the $\Delta t_{P2-Pr}$ to characterise how the lattice response evolves in time after P1. Fig. 4a shows the result of this calculation for $\Phi_{P1} = 26$ μJ/cm$^2$.

The interpretation of the coherent component is simpler as it describes the phonon modes that P2 can excite at each $\Delta t_{P1-P2}$ delay.

Oscillations in the 2, 2.9, and 3.6 THz phonon amplitude are present, as it is clear in Fig. 4b where we present the linecuts along the resonance frequencies. Nodes and antinodes of these oscillations occur as P2 coherently amplifies or suppresses the phonons excited by P1[63,64], and are an indication of these modes' coherence. Interestingly, the 1 THz mode is only very weakly modulated, and instead appears partially suppressed at early delays. When we increase $\Phi_{P1}$ (see Fig. 4c), while the 2.9 and 3.6 THz modes are not significantly affected, a drastic change is observed for the other two modes. As evidenced in Fig. 4d, the 1 THz mode is progressively quenched at early times, until it is completely suppressed at the highest $\Phi_{P1}$, where it starts recovering only after a delay of ~0.8 ps. We associate this behaviour with a saturation of the 1 THz mode excitation, which can be also deduced from the two-pulse results (see Supplementary Information, Note 8). As $\Phi_{P1}$ is increased, the perturbed, photoexcited state of the system cannot support the excitation of additional 1 THz phonons, and therefore the $\Delta(\Delta R/R)$ signal does not contain any signature of this mode at short P1-P2 delays. In addition, the 1 THz phonon mode exhibits a temperature dependent saturation (see Supplementary Information Note 8): at high temperatures, the amplitude of the oscillations are nearly independent of $\Phi_{pump}$. Moreover, we note that the recovery time of the 1 THz amplitude closely mirrors $\tau_{cond.}$. This suggests that the saturation of the 1 THz mode is associated with the depletion of the excitonic condensate, but further studies are required to unveil this connection. These results add important insights on the behaviour of this phonon mode, suggesting a possible different interpretation of its role from what was previously reported[23,27].

Even more important, for what concerns our analysis, is the response of the 2 THz phonon oscillations. As seen in Fig. 2d, this mode provides information on the evolution in the lattice properties. In Fig. 4c we see that this mode is strongly affected at early times, but it still shows a coherent oscillation induced by P2 before reaching a constant value, which signals the return to the weakly perturbed state. A better understanding of the quenching process can be gained by taking finer P1-P2 delay steps and achieving a higher signal-to-noise ratio (Fig. 4e). Here, we observe that the 2 THz mode is promptly shifted to lower frequencies and broadened, while still showing a modulation induced by P2, signalling a residual coherence of the oscillations excited by P1. At intermediate values of $\Phi_{P1}$ (Fig. 4f) a signal indicative of a partial collapse of the EI phase is observed as expected. We therefore conclude that the lattice transitions to its high temperature response[44] on a femtosecond timescale, before relaxing back to the typical room temperature behaviour in ~1.5 ps. This supports our interpretation that, upon strong photoexcitation, the excitonic order can be completely suppressed and recovered on timescales compatible with the presence of strong correlation driven by electronic interactions. We stress that all the results obtained from the incoherent and coherent components of the signal, both with the pump-probe and the three-pulse experiments, are important pieces of evidence that were used to construct our interpretation.

Previous measurements on $Ta_2NiSe_5$ have suggested that structural degrees of freedom play an important role in the formation of the EI phase, raising questions about the nature of the symmetry breaking transition. In particular, it is crucial to determine the relative strength of the electronic and structural contributions to the phase transitions. In other words, we want to determine if the electronic contribution is strong enough that, if we were able to suppress the structural phase transition, we could still obtain the excitonic order, possibly manifesting superfluidic behaviour. In this work, we used a combination of temperature and fluence dependent, ultrafast spectroscopy techniques to address this question and to provide a more complete picture of the femtosecond dynamics of $Ta_2NiSe_5$. In analogy with the case of high temperature superconductors, we find that high-energy electronic transitions track the low-energy physics of the system, allowing us to use the response in the NIR as a marker of the EI order parameter.

Through the study of both the electronic and lattice out-of-equilibrium response, we find evidence that the excitonic long-range order can be transiently melted under strong photoexcitation. Crucially, the timescale of the EI phase suppression supports the hypothesis of an excitonic order driven substantially by electronic interactions.

Nonetheless, our work also showcases the existence of strong electron-phonon interactions in $Ta_2NiSe_5$ and could guide the development of experimental and theoretical efforts aimed at uncovering the nature of the connections between these degrees of freedom. In particular, the relationship between the possible state with short-range order and the behaviour of the 1 and 2 THz phonon modes would be an area of great research interest.

We have also accessed the out-of-equilibrium dynamics of the high-temperature phase of $Ta_2NiSe_5$, believed to host a state of preformed excitons. Further investigations of this insulating phase could shed more light onto the properties of this and other systems characterized by strong electronic interactions. Intriguingly, the discovery of a photoinduced, ultrafast sign reversal and recovery of the material reflectivity also showcases the possibility of using $Ta_2NiSe_5$ as a platform for the development of ultrafast optical switches and more in general for quantum devices based on the control of correlated phases[65].

## Methods

**Sample growth and preparation:** The sample was grown with the procedure outlined in reference [66]. To prepare samples for optical measurements, flakes were exfoliated onto glass coverslips and subsequently encapsulated with a second glass slide while in a nitrogen glovebox.

**Ultrafast spectroscopy measurements**

Pump-probe measurements are carried out using a Nd:YAG laser system (PHAROS, Light Conversion), which delivers a 14.5 W output of 1025 nm light, 200 fs pulse duration, and with a repetition rate of 38 kHz. Part of the output further seeds a commercial OPA (ORPHEUS, Light Conversion). The second harmonic of the Nd:YAG laser, generated by the OPA, is combined with the fundamental laser beam in a non-collinear OPA to generate broadband pulses centered around 1.8 eV, which are compressed to ~12 fs using a wedge prism and chirp mirrors. The compressed beam was split to generate Pump1 and Pump2, whose time delay relative to each other are controlled with mechanical delay stages. The broadband probe pulses (550-1250 nm) are generated from the fundamental of the Nd:YAG laser using a sapphire crystal, and their delay with respect to the pump pulses is controlled with a mechanical delay stage. Pump and probe beams are then combined in a boxcar geometry on the sample. Reflected pump pulses are discarded while the reflected probe pulses are collected with a collimating lens and split using a dichroic mirror with a 960 nm cutoff to separate the visible and near-infrared components of the beam, which are then detected by an InGaAs and Si photodiode array, respectively. Both arrays are synchronised and read-out at the repetition rate of the laser, using a chopping scheme of 38/4 kHz for Pump1 and 38/8 kHz for Pump2. For further details, see reference [67].

## Data availability

The data underlying all figures in the main text are publicly available at https://doi.org/10.17863/CAM.63831

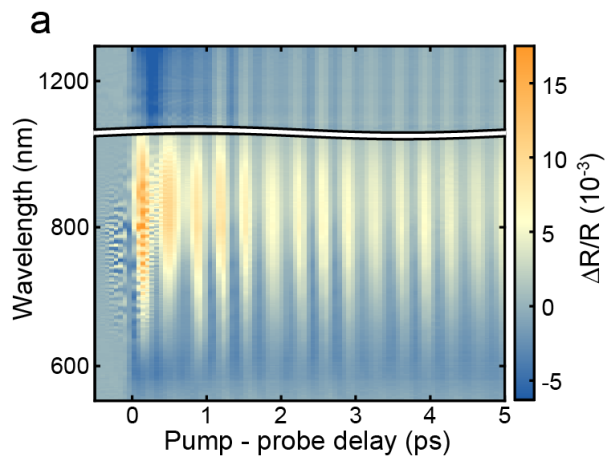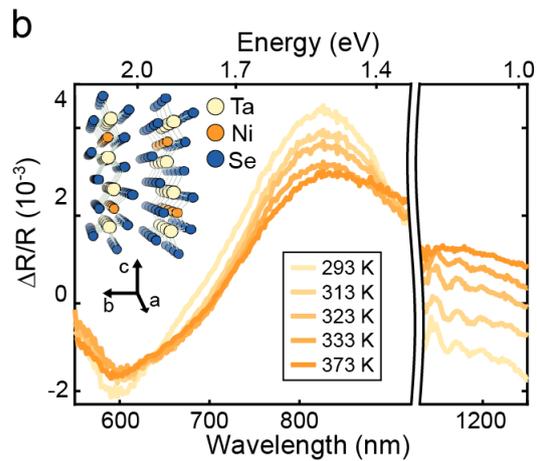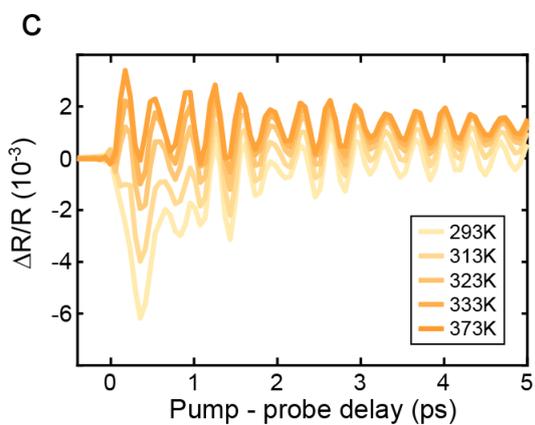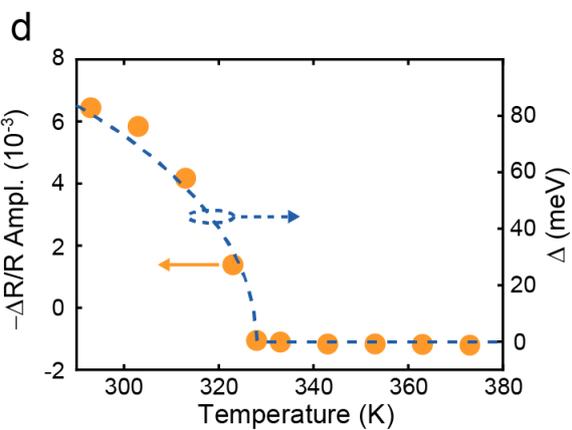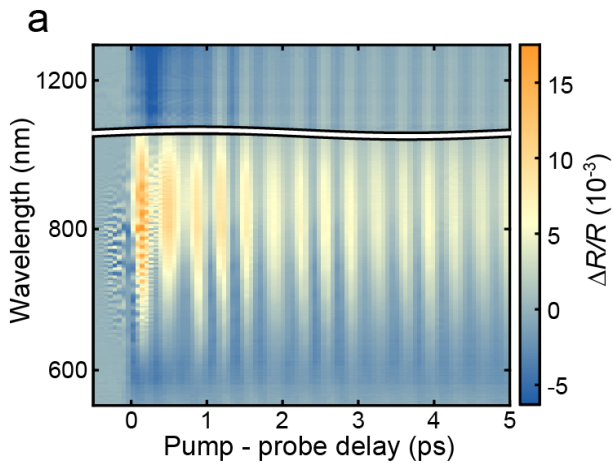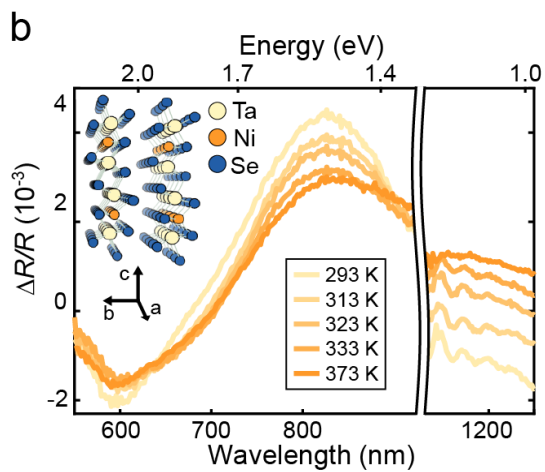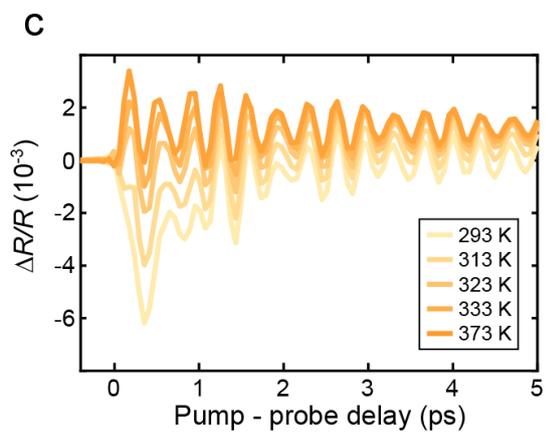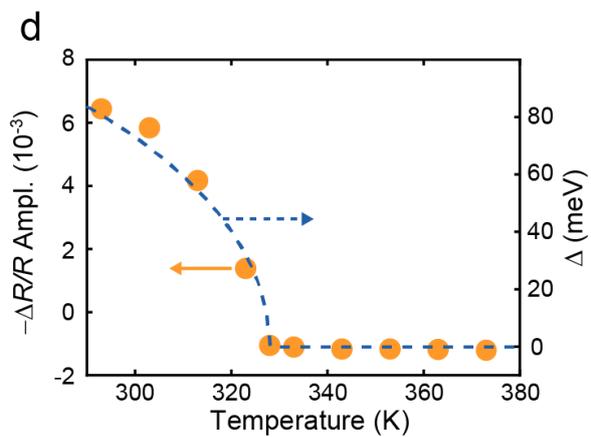

Fig. 1. Temperature dependence of the transient reflectivity signal of Ta$_2$NiSe$_5$. (a) $\Delta R/R$ signal measured at room temperature as a function of pump – probe delay and probe wavelength for $\Phi_{pump}$ = 50 µJ/cm$^2$. (b) Spectra of $\Delta R/R$ averaged between 1 and 2 ps for a series of temperatures and $\Phi_{pump}$ = 50 µJ/cm$^2$. The range between 930 and 1130 nm is removed in (a) and (b) to eliminate the noise due to residual spectral components of the laser pulses used to generate the probe beam. (c) $\Delta R/R$ kinetic at 1200 nm corresponding to the data in (b). (d) Temperature dependence of the signal amplitude (orange dots, left y-axis) and of a BCS-like order parameter calculated using[37] $\Delta(T)/\Delta(0) \approx 1.74(1 - T/T_c)^{0.5}$, valid for $T \approx T_c$ (blue dashed line, right y-axis). The amplitudes are extracted from a fit of the incoherent component of the signal (See Supplementary Information Note 2 for more details). In the inset of figure (b) we show the crystal structure of TaNiSe (figure generate with the software Vesta[68]).

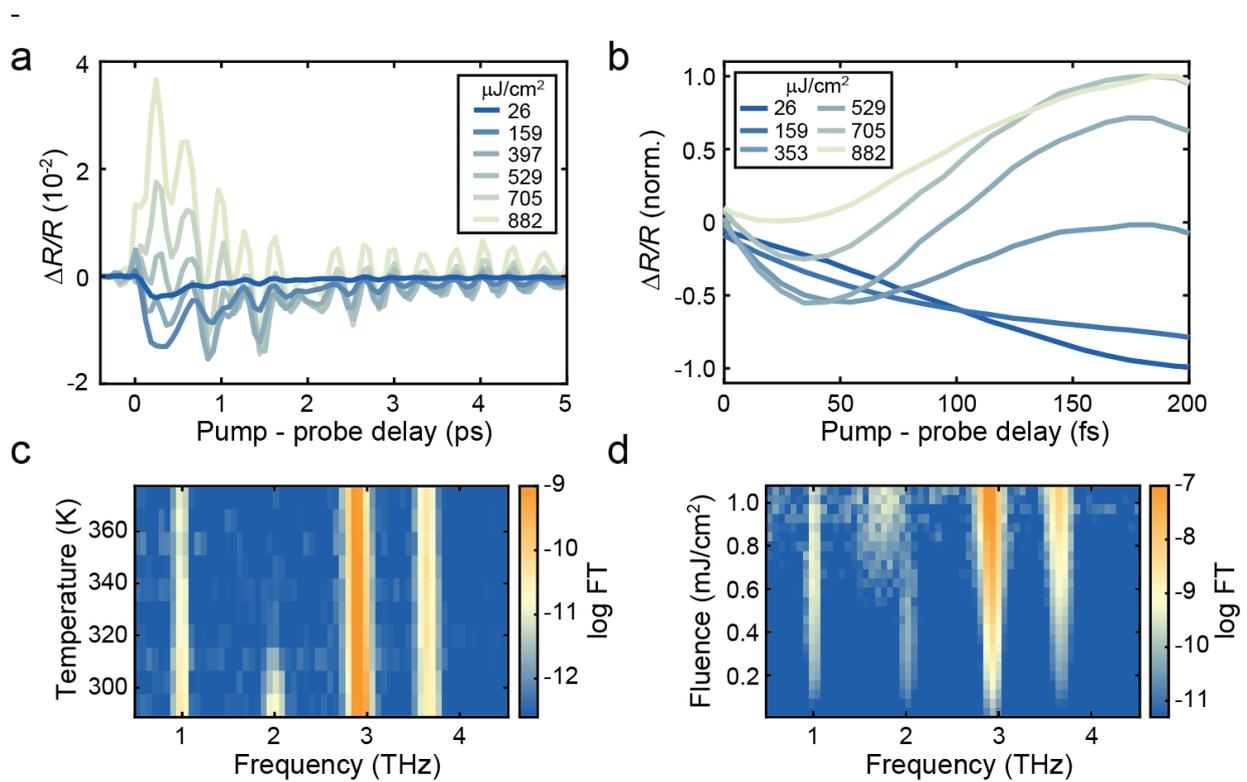

Fig. 2. Photoinduced suppression of the excitonic order. (a) $\Delta R/R$ kinetic at 1200 nm for different values of $\Phi_{pump}$. (b) Normalised signal collected with smaller time steps to better resolve the behaviour at early times. Data is cut after the coherent artefact for clarity (see Supplementary Information, Note 4). (c,d) Natural logarithm of the Fourier transform (FT) of the oscillatory component of $\Delta R/R$ as a function of temperature (c) and $\Phi_{pump}$ (d).

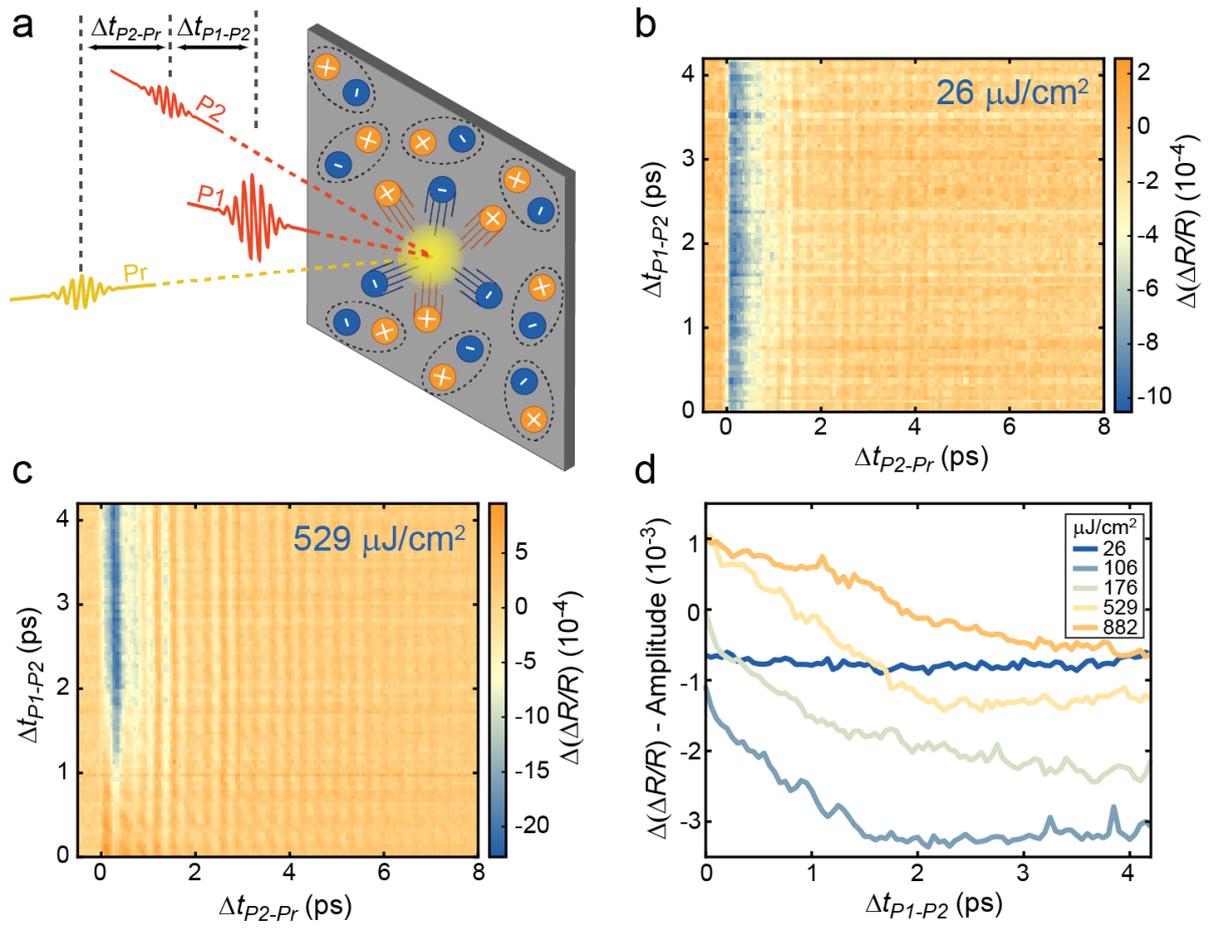

Fig. 3. Recovery of the EI phase measured with multi-pulse ultrafast spectroscopy. (a) Schematic of the multi-pulse experiment. (b,c) $\Delta(\Delta R/R)$ signal at 1200 nm as a function of the P2 – Pr and P1 – P2 delays for $\Phi_{P1} = 26$ µJ/cm$^2$ (b) and $\Phi_{P1} = 529$ µJ/cm$^2$ (c). (d) Evolution of the $\Delta(\Delta R/R)$ amplitude as a function of the P1 – P2 delay for different values of $\Phi_{P1}$.

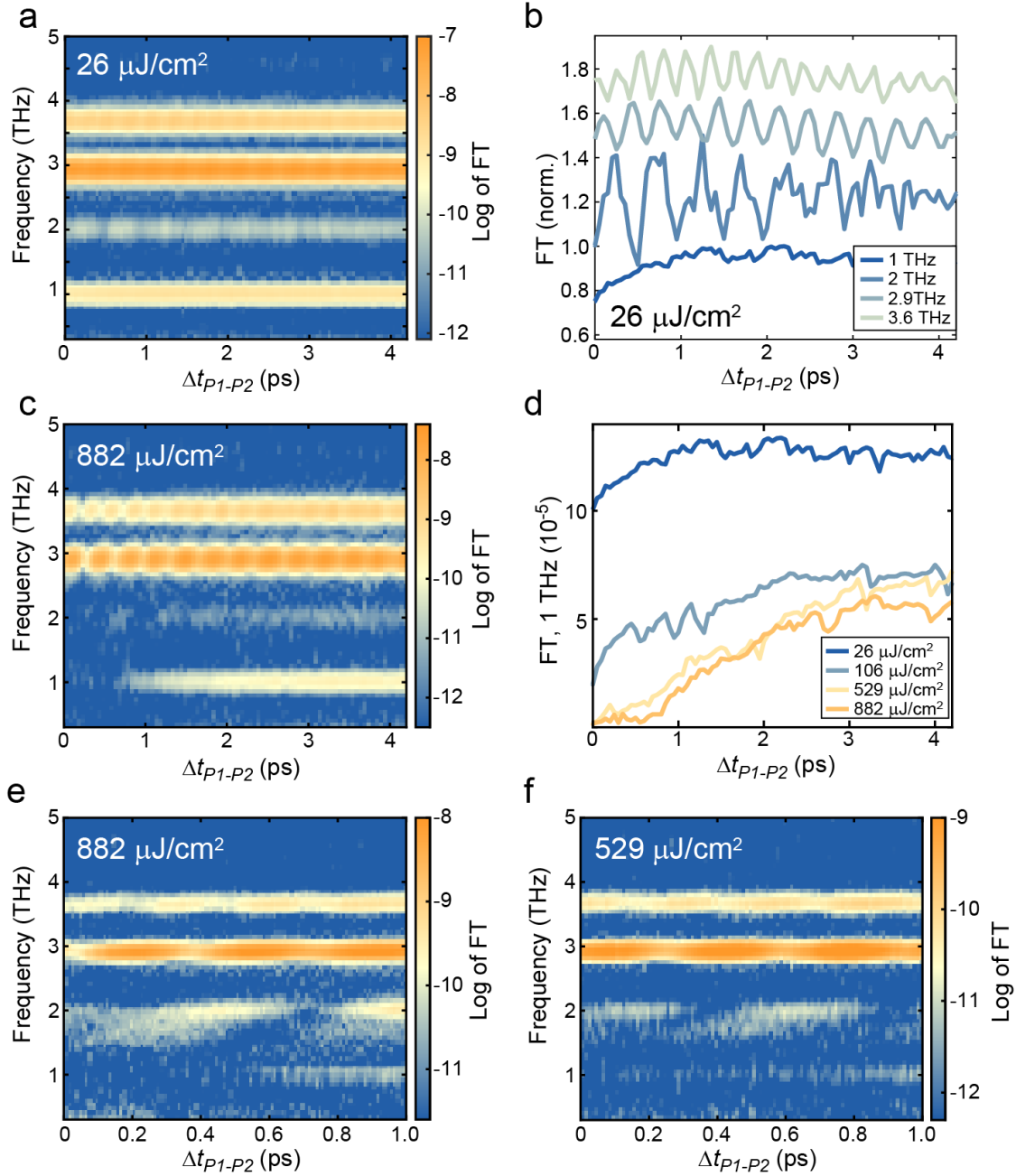

Fig. 4. Effect of $\Phi_{P1}$ on the coherent phonon oscillations excited in the system. (a,c) Natural logarithm of the FT of $\Delta(\Delta R/R)$ oscillatory component at 1200 nm as a function of the P1 – P2 delay for $\Phi_{P1} = 26\ \mu J/cm^2$ (a) and $\Phi_{P1} = 882\ \mu J/cm^2$ (c). (b) Normalized linecuts at 1, 2, 2.9, and 3.6 THz of the data in (a). (d) FT amplitude for the 1 THz resonance as a function of the P1 – P2 delay for a series of $\Phi_{P1}$ values. (e,f) Data analogous to (a,c) but collected with higher temporal resolution for $\Phi_{P1} = 882\ \mu J/cm^2$ (e) and $\Phi_{P1} = 529\ \mu J/cm^2$ (f).

## Acknowledgements

The authors would like to thank Alexander Boris for kindly sharing data of the equilibrium optical conductivity of $Ta_2NiSe_5$. We would like to thank Yuta Murakami, Denis Golež, Selene Mor, Daniele Fausti, Vikas Arora, and Edoardo Baldini for many fruitful discussions. We are also grateful to the Engineering and Physical Science Research Council (EPSRC) and the Winton Programme for the Physics of Sustainability for funding. We acknowledge the financial support from the Department of Science and Technology (DST), India [Grant No. SR/WOS-A/PM-33/2018 (G)] and IISER Pune for providing the facilities for crystal growth and characterization. We acknowledge funding from the European Research Council (ERC) under the European Union's Horizon 2020 research and innovation programme (Grant Agreement 758826). We thank the Department of Science and Technology, India for support under Nanoemission and Year of Science Professorship.

## Author Contributions

HMB and PA performed measurements and analysed data. PT, AS, and LH worked on the growth and characterization of the material. PA, AKS and AR conceived the experiments. HMB, PA, and AR wrote the paper with input from all authors.

## Competing Interests

The authors declare no competing interests.